\documentclass[hyper,12pt]{JHEP3}
\usepackage{amsmath,amsfonts}
\usepackage{graphicx}

\title{Asymptotically free ${\cal N}=2$ theories and irregular conformal blocks}

\author{Davide Gaiotto \\
School of Natural Sciences, Institute for Advanced Study, \\
Princeton, NJ 08540, USA
\\
{\tt dgaiotto@ias.edu} }

\preprint{ }

\abstract{A surprising connection between $N=2$ gauge theory instanton partition functions and conformal blocks has been recently proposed. We illustrate through simple examples the generalization to asymptotically free $N=2$ gauge theories
}

\begin{document}
\section{Introduction}

In \cite{Alday:2009aq} evidence was presented for a direct correspondence between the instanton partition functions \cite{Nekrasov:2002qd,Nekrasov:2003rj} of
${\cal N}=2$ superconformal quivers of $SU(2)$ gauge groups (``$A_1$ superconformal quivers'' in short)
and Virasoro conformal blocks.
At the root of the correspondence lies a specific realization \cite{Witten:1997sc,Gaiotto:2009hg,Gaiotto:2009we} of the $A_1$ superconformal quivers as the twisted compactification of a six dimensional theory: the $A_1$ $(2,0)$ six dimensional SCFT. Asymptotically free quivers of $SU(2)$ gauge groups ($A_1$ quivers) admit a very similar realization \cite{Witten:1997sc,Gaiotto:2009hg}. The only new ingredient is a larger class of codimension two defects in the six dimensional theory.

All theories in the $A_1$ class are associated to a meromorphic quadratic differential $\phi_2$ on a Riemann surface, which is allowed to have poles at the location of the defects. The poles are always of degree $2$ or smaller  for $A_1$ superconformal quivers. (The terminology ``regular'', or ``tame'' could be used for these singularities). General theories in the $A_1$ class are associated to a quadratic differential with poles of higher order.(The terminology ``irregular'', or ``wild'' could be used for these singularities). This has a simple physical motivation.  These asymptotically free theories are obtained from superconformal theories by tuning some mass parameter to be very large, while adjusting the marginal gauge coupling in the UV so that the running coupling in the IR remains finite.  From a six-dimensional perspective, the limiting procedure brings two or more standard punctures together to produce a single puncture with a larger degree of divergence for $\phi_2$.

It would be reasonable to carry over the same limiting procedure to the instanton partition function, and then the corresponding  conformal block, in order
to extend the results of \cite{Alday:2009aq} to asymptoticlly free theories.
Instead, we will follow here an instructive shortcut.
We will define directly some exotic states in the CFT which correspond to the irregular singularities of $\phi_2(z)$. ``Irregular'' conformal blocks with insertions of such states at punctures will reproduce the instanton partition functions of the asymptotically free theories. We refer to \cite{Moore:1990mg}, \cite{Moore:1990cn}, \cite{Miwa:1980tu} for previous attempts to define such type of states in the $c=1$ theory.

We will use the standard dictionary of \cite{Alday:2009aq}: set a scale by $\epsilon_1 \epsilon_2=1$ and identify $\epsilon_1 = b$, $Q=b + \frac{1}{b}$, $c=1+6 Q^2$. The expressions for the quadratic differential are taken from \cite{Gaiotto:2009hg}, possibly with trivial redefinitions/rescalings of the parameters.
The form of the Seiberg-Witten curves $x^2 = \phi_2(z)$ may be unfamiliar to the reader,
as it differs in form from the original expression \cite{Seiberg:1994rs,Seiberg:1994aj}. This form of the curve follows naturally from the brane construction in \cite{Witten:1997sc}, and the relation to the spectral curve of a Hitchin system.
The conventions about mass parameters in the instanton partition function are such that $m \to -m$ exchanges fundamental and antifundamental matter contributions. Masses are shifted by $Q/2$ with respect to the convention for which $m \to Q-m$ exchanges fundamental and antifundamental matter contributions

\section{$SU(2)$ $N_f=0$}\label{sec:nf0}
This theory is associated to a sphere with two punctures, such that the quadratic differential
$\phi_2$ has a pole of degree $3$ at each puncture. If the punctures are set at $z=0, \infty$
we can take
\begin{equation}
\phi_2 = \frac{\Lambda^2}{z^3} + \frac{2 u}{z^2} + \frac{\Lambda^2}{z}.
\end{equation}
Here $\Lambda$ is fixed, and coincides with the scale of the $SU(2)$ theory, while $u$ parameterizes the Coulomb branch.

$\phi_2$ has been identified in \cite{Alday:2009aq} with the semiclassical limit of the energy momentum tensor $T(z)$.
We would like to consider a two dimensional ``conformal block'' with two special punctures on the sphere, i.e. the inner product
\begin{equation}
\langle \Delta,\Lambda^2| \Delta,\Lambda^2\rangle.
\end{equation}
The state $| \Delta,\Lambda^2\rangle$ should live in the Verma module of a
highest weight state of conformal dimension $\Delta = \frac{Q^2}{4} - a^2$.
We hope to identify $\pm a$ with the eigenvalues of the vector multiplet scalar in the instanton partition function.
To reproduce the singularity of $\phi_2(z)$ at $z=0$ we are led to the requirements
\begin{equation}
L_1 | \Delta,\Lambda^2\rangle = \Lambda^2  | \Delta,\Lambda^2\rangle \qquad \qquad  L_2 | \Delta,\Lambda^2\rangle  =0.
\end{equation}
Notice that the Virasoro commutation relations are sufficient to imply then $L_n  | \Delta,\Lambda^2\rangle=0$ for all $n>2$.

We aim to define $| \Delta,\Lambda^2\rangle$ as a (possibly formal) power series in $\Lambda^2$,
i.e,
\begin{equation}
| \Delta,\Lambda^2\rangle = v_0 + \Lambda^2 v_1 + \Lambda^4 v_2 + \cdots
\end{equation}
Here $v_0$ is the highest weight vector $|\Delta\rangle$ and $v_n$ is a level $n$ descendant such that $L_1 v_n = \Lambda^2 v_{n-1}$ and $L_2 v_n=0$. It is not fully clear to us why such vectors should exist.
Surprisingly, we found that these equations can be recursively, and uniquely solved to as high a level $n$ as we cared to check. (level $8$)

The inner product
\begin{equation}
\langle \Delta,\Lambda^2| \Delta,\Lambda^2\rangle = \sum \Lambda^{4n} |v_n|^2
\end{equation}
coincides order by order (again, we only checked up to level $8$) with the instanton partition function for $SU(2)$ $N_f=0$, with the simple identification of instanton factor as $q=\Lambda^4$

For reference, we report here the first few $v_n$
\begin{align}
v_0 &= |\Delta\rangle \notag \\
v_1 &= \frac{1}{2\Delta} L_{-1}|\Delta\rangle \notag \\
v_2 &=\frac{1}{4 \Delta \left(2 c
   \Delta+c+16 \Delta^2-10 \Delta \right)} ((c+8 \Delta) L_{-1}^2-12 \Delta L_{-2})|\Delta\rangle \notag \\
v_3 &=\frac{1}{24 \Delta (2 + c - 7 \Delta + c \Delta + 3 \Delta^2) (c + 2 c \Delta + 2 \Delta (-5 + 8 \Delta))} \notag \\&(12 \Delta (-3 - c + 7 \Delta) L_{-3} -
 12 (c + 3 c \Delta + \Delta (-7 + 9 \Delta))L_{-2} L_{-1}+\notag \\ &+(c^2 + c (8 + 11 \Delta) + 2 \Delta (-13 + 12 \Delta))L_{-1}^3 ) |\Delta\rangle.
\end{align}

\section{$SU(2)$ $N_f=1$}\label{sec:nf1}
This theory is associated to a sphere with two punctures, such that the quadratic differential
$\phi_2$ has a pole of degree $3$ at a puncture (say at $z=0$) and a pole of degree $4$ at the other puncture (say at $z=\infty$).
We can make some convenient choices
\begin{equation}
\phi_2 = \frac{\Lambda^2}{2 z^3} + \frac{2 u}{z^2} - \frac{2\Lambda m}{z} - \Lambda^2  .
\end{equation}
Here $\Lambda$ is fixed, and coincides with the scale of the $SU(2)$ theory. $m$ is also fixed, and coincides with the mass of the single flavor hypermultiplet. $u$ parameterizes the Coulomb branch. The minus signs and the $\frac{1}{2}$ factor are introduced to simplify later expressions.

We will consider the following inner product:
\begin{equation}
\langle \Delta,\Lambda,m| \Delta,\Lambda^2/2\rangle.
\end{equation}
Both states  should live in the Verma module of the
highest weight state of conformal dimension $\Delta = \frac{Q^2}{4} - a^2$.  $| \Delta,\Lambda^2/2\rangle$ is the same state as in the previous section, with a trivial redefinition of $\Lambda^2$ . We require $| \Delta,\Lambda,m\rangle$ to satisfy
\begin{equation}
L_2 | \Delta,\Lambda,m\rangle = -\Lambda^2  | \Delta,\Lambda,m \rangle \qquad \qquad  L_1 | \Delta,\Lambda,m\rangle =- 2 m \Lambda  | \Delta,\Lambda,m\rangle .
\end{equation}
Notice that the Virasoro commutation relations are again sufficient to imply $L_n  | \Delta,\Lambda,m\rangle=0$ for all $n>2$.

We aim to define $| \Delta,\Lambda,m\rangle$ as a (possibly formal) power series in $\Lambda$,
i.e,
\begin{equation}
| \Delta,\Lambda,m\rangle = w_0 + \Lambda w_1 + \Lambda^2 w_2 + \cdots
\end{equation}
Here $w_0$ is the highest weight vector $|\Delta\rangle$ and $w_n$ is a level $n$ descendant such that $L_1 w_n = - 2 m  w_{n-1}$ and $L_2 w_n=- w_{n-2} $.
Again, it is not fully clear to us why such vectors should exist, but we checked their existence and unicity for the first few $n$.

The inner product
\begin{equation}
\langle \Delta,\Lambda,m| \Delta,\Lambda^2/2\rangle = \sum \Lambda^{3n} 2^{-n} \langle w_n| v_n \rangle
\end{equation}
coincides order by order (again, we only checked the first few
levels) with the instanton partition function for $SU(2)$
$N_f=1$, with $q=\Lambda^3$ and with mass parameter $m$.

As a reference, we report here the first few $w_n$
\begin{align}
w_0 &= |\Delta\rangle \notag \\
w_1 &=- \frac{m}{\Delta} L_{-1}|\Delta\rangle \notag \\
w_2 &=\frac{1}{\Delta (c + 2 c \Delta + 2 \Delta (-5 + 8 \Delta))} ((c m^2 + \Delta (3 + 8 m^2)) L_{-1}^2-2 \Delta (1 + 2 \Delta + 6 m^2)L_{-2})|\Delta\rangle.
\end{align}

\section{$SU(2)$ $N_f=2$, first realization}\label{sec:nf2}
This theory has two distinct realizations. The first realization  is associated to a sphere with two punctures, such that the quadratic differential
$\phi_2$ has a pole of degree $4$ at both punctures (say at $z=0,\infty$).
We can take some convenient choices
\begin{equation}
\phi_2 = -\frac{\Lambda^2}{ z^4} - \frac{2\Lambda m_1}{z^3} + \frac{2 u}{z^2} - \frac{2\Lambda m_2}{z} - \Lambda^2.
\end{equation}
Here $\Lambda$ is fixed, and coincides with the scale of the $SU(2)$ theory. $m_{1,2}$ are also fixed, and coincide with the masses of the two flavor hypermultiplets. $u$ parameterizes the Coulomb branch.

We will consider the following inner product:
\begin{equation}
\langle \Delta,\Lambda,m_2| \Delta,\Lambda,m_1\rangle.
\end{equation}
Here we meet a spurious ``$U(1)$ factor''. The inner product coincides order by order (as usual, we only checked the first few levels) with the instanton partition function for $SU(2)$ $N_f=2$, with $q=4\Lambda^2$,
multiplied by an overall factor which depends on $\Lambda$ only.
The overall factor has a power expansion which we conjecture to coincide with $\exp 2 \Lambda^2$.
The appearance of this spurious factor is hardly a surprise. The spurious factor for $N_f=4$ was
some $(1-q)^{2 \tilde m_0( Q-\tilde m_1)}$, where $\tilde m_{0,1}$ are appropriate combinations of the mass parameters. In the asymptotically free limit, $q\to0$ as $(\Lambda/M)^{4-N_f}$, where $M$ is the mass scale sent to infinity.
The spurious factor goes to $1$ if $N_f<2$, but can have a finite limit otherwise. In this realization of the $N_f=2$ theory, both $m_{0,1}$ scale as $M$, leading to a finite limit.

\section{$SU(2)$ $N_f=2$, second realization}\label{sec:nf2bis}
There is a second way to realize the same four dimensional theory from a distinct six dimensional setup.
The setup contains two regular punctures (say at $z=0,1$) and a puncture with a pole of rank $3$
(say at $z=\infty$).
\begin{equation}
\phi_2 = -\frac{m_+^2}{z^2}-\frac{m_-^2}{(z-1)^2} + \frac{2 \tilde u}{z(z-1)} + \frac{\Lambda^2}{z} .
\end{equation}

We will consider a three point function, involving the highest weights of dimensions $
\Delta_{\pm} = \frac{Q^2}{4} - m_{\pm}^2$ and $| \Delta,\Lambda^2\rangle$. The masses $m_{\pm}$ are such that the fundamental masses are $m_1 = m_+ + m_-$ and $m_2 = m_+ - m_-$The calculation is straightforward,
and reproduces the instanton partition function right away, with no overall spurious factors and $q=\Lambda^2$

\section{$SU(2)$ $N_f=3$}\label{sec:nf3}
Finally, to get a theory with three flavors we have to consider
a setup which contains two regular punctures (say at $z=0,1$) and a puncture with a pole of rank $4$
(say at $z=\infty$).
\begin{equation}
\phi_2 = -\frac{m_+^2}{z^2}-\frac{m_-^2}{(z-1)^2} + \frac{2 \tilde u}{z(z-1)} - \frac{2 m_3 \Lambda}{z} - \Lambda^2.
\end{equation}

We will consider a three point function, involving the highest weights of dimensions $
\Delta_{\pm} = \frac{Q^2}{4} - m_{\pm}^2$ and $| \Delta,\Lambda,m_3\rangle$. The calculation is straightforward,
and reproduces the instanton partition function, with $q=- 2 \Lambda$, multiplied by what appears to be
$\exp(1/b + b - 2 m_1)$. Again,  $m_1 = m_+ + m_-$ and $m_2 = m_+ - m_-$. The spurious factor has a simple interpretation as the appropriate asymptotically free limit of the $N_f=4$ spurious factor.

\section{Concluding remarks}
It appears that the special punctures required to realize the
instanton partition function of asymptotically free theories as
2d CFT ``conformal blocks'' can be defined in a straightforward
way, requiring them to be eigenvectors of a set of Virasoro
generators which is in one-to-one correspondence with the
coefficients of $\phi_2$ which are fixed in gauge theory.
Although the examples we provide only cover singularities up to
$z^{-4}$, higher degrees of singularity should work in a
similar fashion. Notice that the proper formulation for the
boundary conditions of $\phi_2$ at an irregular singularity
involves fixing the singular part of $\sqrt{\phi_2}$. If the
degree of singularity of $\phi_2$ is $2N$ (or $2N-1$), the
coefficients in the Laurent polynomial for $\phi_2$ from
$z^{-2N}$ to $z^{-N-1}$ are fixed. The corresponding Virasoro
generators $L_n$ with $n$ from $2N-2$ to $N-1$ can act as a
multiple of the identity when acting on a state killed by
$L_{n'}$ with $n'>2N-2$, without violating the Virasoro
algebra. We conjecture that such a state always exist (at least
as a formal, level-by-level sum) and is unique.

\section*{Acknowledgments}
The author has benefited from discussions with Y.Tachikawa and
L.F.Alday. We also want to thank Y.Tachikawa for sharing with
us some powerful Mathematica notebooks devoted to instanton
computations.  D.G. is supported in part by the DOE grant
DE-FG02-90ER40542. D.G. is supported in part by the Roger
Dashen membership in the Institute for Advanced Study.

\bibliography{Liouville}{}
\end{document}